\documentstyle[twocolumn,aps,psfig]{revtex}
\draft
\def\be{\begin{equation}}
\def\ee{\end{equation}}
\begin{document}
\twocolumn[\hsize\textwidth\columnwidth\hsize\csname 
@twocolumnfalse\endcsname
\title{Superconductivity and  Quantum Spin Disorder in Cuprates}
\author{Moshe Havilio\cite{email1} 
and Assa Auerbach\cite{email2}}
\address{Physics Department, Technion, Haifa 32000, Israel}
\date{\today}
\maketitle
\begin{abstract}
A fundamental connection between superconductivity 
and  quantum spin fluctuations in underdoped cuprates, is revealed. 
A variational calculation shows that
{\em  Cooper pair hopping} strongly reduces the local magnetization $m_0$.
This effect pertains to recent 
neutron scattering and  muon spin rotation measurements 
in which $m_0$ varies weakly with hole doping in the
poorly conducting regime,  but drops precipitously above the onset
of superconductivity.
\end{abstract}
\pacs{74.20.Mn, 75.10.Jm, 71.10.Fd}
\vskip1pc]
\narrowtext

When holes are introduced into the copper oxide planes of high T$_c$ cuprates,
spin and charge correlations change dramatically. 
The  {\em local} magnetization $m_0$, measured by  $\mu$SR  
\cite{Niedermayer} and elastic neutron scattering\cite{Wakimoto} on e.g. $\mbox{La}_{2-x}\mbox{Sr}_x \mbox{CuO}_4$,
reveals a qualitative difference 
between the insulating and superconducting phases: $m_0$ is rather insensitive to doping in the poorly conducting
regime 
$0\le x\le 0.06$, but drops precipitously above the onset  of  superconductivity at $x > 0.06$, 
becoming undetectable at optimal doping  $x\approx 0.15$. Theoretically, 
holes can cause {\em dilution} and 
{\em frustration} in  the Heisenberg antiferromagnet, which create spin textures: 
either random (``spin glass'') or with ordering wavevector away from $(\pi,\pi)$ 
(sometimes called ``stripes'')\cite{IC}. However, the apparent reduction of local magnetization 
by the onset of superonductivity, is a novel and poorly understood effect. 
Theory must go beyond
purely magnetic models, and
involve the superconducting degrees
of freedom.  

We find that this problem is amenable to a variational approach, using  hole-doped Resonating Valence Bonds (RVB)  
states, originally suggested by Anderson for the spin correlations
of high $T_c$ cuprates\cite{PWA,LDA,BW}. 

These RVB states are excellent trial states for 
doped Mott insulators, with large Hubbard repulsion $U$:\newline
(i) Configurations with doubly occupied sites are excluded.\newline
(ii)  Marshall's sign criterion for
the magnetic energy\cite{Marshall} is satisfied, and Heisenberg
antiferromagnetism at  zero doping is accurately recovered.\newline
(iii) For doped systems, spin and charge correlations are parameterized independently,
without explicit spin nor gauge symmetry breaking.

These are important advantages over commonly
used Spin Density Wave, Hartree-Fock and BCS wavefunctions for
the antiferromagnet, metal  and superconducting phases respectively. 
RVB states  permit an {\em unbiased} determination of 
ground state spin and charge correlations appropriate for the cuprates.

A phenomenological low energy effective Hamiltonian is used,
with two major components:
Heisenberg interaction for spins, 
and single or Cooper pair hoppingng kinetic energy for fermion holes.

Our key results are as follows:\newline
(i)  For the magnetic energy alone,  the local magnetization $m_0$  is 
{\em weakly dependent} on doping concentration. This holds
independently of inter-hole correlations for either
randomly localized or extended states.\newline
(ii)  In contrast to (i), $m_0$ is strongly reduced by the kinetic
energy of {\em Cooper pair hopping}, which correlates the reduction of $m_0$
with the  rise of
superconducting stiffness, and  hence\cite{emkivsfs} the
transition temperature $T_c$.

Our  results agree with the experimentally reported correlation between $m_0$ and  $T_c$\cite{Niedermayer,Wakimoto}. 
This relation appears to be  {\em weakly} dependent on the precise
hole density.
A brief  discussion concludes the paper.

\begin{figure}
\input epsf
\centerline{ \epsfysize 3.0cm
\epsfbox{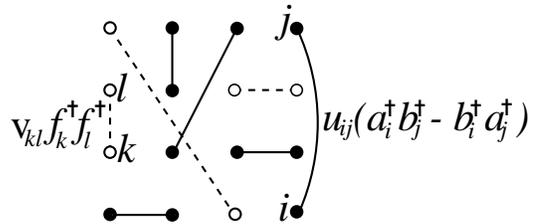}}
\caption{A bond configuration in the doped RVB states $\Psi[u,v]$.
Solid (empty) circles represent spins (holes) with bond correlations $u_{ij}$ ($v_{kl}$). }
\label{rvbfig}  	 	 
\end{figure}

{\em The Wavefunctions:}
The hole-doped  
RVB states are compactly defined by
\begin{eqnarray}
 \Psi [u,v;x]  &=& {\cal P}_G(x) \bar{\psi}[u,v]\nonumber\\
  \bar{\psi} [u,v]&\equiv&
\exp\left(
\sum_{ij}\left( v_{ij} f^\dagger_i f^\dagger_j +u_{ij}(a^\dagger_i b^\dagger_j-b^\dagger_i a^\dagger_j \right)\right)|0\rangle,\nonumber\\
{\cal P}_G(x)&\equiv&  \delta\left(  \sum_i n^f_i  -xL^2 \right)\prod_i \delta\left(n^a_i+n^b_i+n^f_i-1 \right).
\label{states2}
\end{eqnarray}
Where $a^\dagger_i,b^\dagger_i$ and  $f^\dagger_i$ are Schwinger bosons and hole fermions respectively\cite{Book},
$i=1,\ldots L^2$ is a site index on the square lattice, and ${\cal P}_G(x)$
is the Gutzwiller projection onto states with no double occupancies.
As a result of the projection, $\Psi$ can be written as a sum over
bond configuations of singlets and hole pairs which cover the lattice as depicted in Fig. \ref{rvbfig}.

$u({\bf r}_{ij})$ and $v({\bf r}_{ij})$ are {\em independent} 
spin and hole bond parameters respectively. $u_{ij}\ge 0$  connects  $i$ on 
sublattice  $A$ to  $j\in B$ respectively, which ensures Marshall's sign.

The expectation value  of an  observable $O$ is computed by a
sum over  loop
coverings:
\begin{eqnarray}
\langle O \rangle  &=&
\sum_{\gamma,\Lambda_\gamma}
W_{\Lambda,\gamma}  O_{\Lambda,\gamma} ,\nonumber\\
W_{\Lambda,\gamma} &=&{1\over \langle \Psi[u,v]| \Psi[u,v]\rangle} {\det_{ij\in \gamma}}^2|v_{ij}|
\prod_{\lambda \in \Lambda_\gamma}\left(2 \prod_{(i, j)\in \lambda} 
u_{ij}\right) .
\label{loops}
\end{eqnarray}
 $\Lambda_\gamma$ denotes a list of  directed loops 
$\{\lambda_\alpha\}$ which cover the  lattice except for  subset $\gamma$
of $xL^2$ 
hole sites.
$W$ are positive
Boltzmann weights, with which the Monte Carlo Metropolis step is defined following Refs. \cite{LDA,BW}.
Ergodicity and convergence of our program was checked
against precise transfer matrix results\cite{mypaper}.

The Gutzwiller Approximation (GA)  
amounts to dropping the projector ${\cal P}(x)$ in state (\ref{states2}) and setting $\Psi\to \bar{\psi}$,
after adjusting the overall normalization 
of  $u$ and $v$ to satisfy the global
constraints  $\langle n^a_i+n^b_i\rangle =(1-x)$,
and $\langle  n^f_i \rangle=x$.  $\bar{\psi}$ is a Fock state of decoupled spins and holes,
with easily computable correlations\cite{Raykin,havilio-thesis}.

For the spin correlations  we use power law decaying functions
$u_p (r)  \equiv 1/r^p $. The single variational parameter $p$ determines the long range spin correlations and
local magnetization\cite{comm-stripes}. 
(Other forms for $u(r)$, with qualitatively similar results, will be described elsewhere\cite{havilio-thesis}).

We discuss four cases of inter-hole correlations:
\begin{equation}
\begin{array}{llll}
v_{ins}^{\gamma}({\bf r}_{ij})&=& \cases{1& $( i,j) \in \gamma$\cr
0& $( i,j) \notin \gamma$}\\
v_{met}({\bf r}) & =& 1/L^2\sum_{{\bf k}\in \Sigma} e^{-i{\bf k}\cdot{\bf r}} \\
v_{\alpha}({\bf r}) &=&\sum_{\hat{\eta}} c_\alpha ({\hat{\eta}})  \delta_{{\bf r}, \hat{\eta}},~~~\alpha=s,d  \\
\end{array}
\label{uv}
\end{equation} 
where $\hat{\eta}$ are nearest neighbor vectors on the square lattice,  $c_s=1$ and $c_d=\hat{\eta}_x^2-\hat{\eta}_y^2$.

$v_{ins}^{\gamma}$ puts the $xL^2$
holes on random sites. This state describes an insulator
with  disordered localized charges.

$v_{met}$ has filled Fermi pockets $\Sigma$, containing $xL^2$ occupied ${\bf k}$-states centered around
 $(\pm \pi/2, \pm \pi/2)$ in the  Brillouin zone \cite{minopi}.
It describes  weakly interacting holes  in a ``metallic'' state.
In real space, $v_{met}({\bf r}_{ij})$  decays slowly as $\sim r^{-3/2}$.  
Correlations in this state were previously computed by Bonesteel and Wilkins\cite{BW}. 

$v_s$ and $v_d$ describe tightly bound hole pairs in relative $s$ and $d$-wave 
symmetry respectively.  

\begin{figure}
\input epsf
\centerline{ \epsfysize 7.0cm
\epsfbox{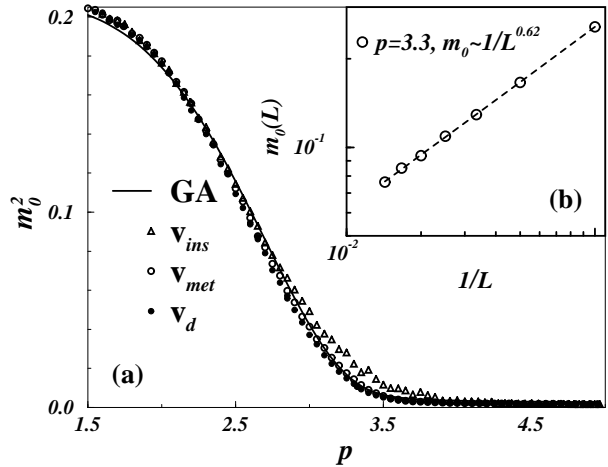}}
\caption{(a) The local magnetization squared of doped RVB wavefunctions
$\Psi[u_p,v]$ versus the variational
power $p$, defined by the bond parameters $u_p(r)=1/r^p$.   Lattice size is
 $40\times 40$, and hole concentration is 10\%.  Results are agree well with
the Gutzwiller approximation (solid line).
The hole bond parameters
$v$ are defined in Eq.\protect{\ref{uv}}. Note that
$m^2_0$ is weakly dependent on $v$; the data for $v=v_s$, overlaps that of $v_d$.
(b) Finite size scaling of $m_0(L)$ for $p=3.3$ which indicates vanishing local magnetization 
at $L\to \infty$. }
\label{sppvsp}  	 	 
\end{figure}

{\em Order parameters:}
The local magnetization $m_0(L)$ is defined by
$m_0^2=1/L^2 \sum_j \langle {\bf S}_i {\bf S}_j\rangle_p e^{-i (\pi,\pi) {\bf r}_{ij} }$.
In Fig.\ref{sppvsp}(a),
$m^2_0(p)$ for $\Psi[u_p,v;x=0.1]$ is plotted 
 for various choices of $v$.
The GA (solid line) seems to work well for $m_0(p)$. 
Finite size scaling in Fig.\ref{sppvsp}(b) indicates vanishing long range order 
$m_0\to 0$ at $p_c= 3.3$, which  lowers the  bound  given previously by Ref. \cite{LDA}:  at $p_c\le  5$.
The GA approximation at $L\to \infty$ suggests that  $p_c\le  3$\cite{havilio-thesis}.
 
The superconducting singlet order parameters are 
\begin{eqnarray}
\Delta^{s,d}_i &=&\sum_{\hat{\eta}}c_{s,d}(\hat{\eta} ) 
 \Delta_{i, i+\hat{\eta}}\nonumber\\
\Delta_{ij} &=&  f^\dagger_i f^\dagger_{j} (a_i b_{j}-b_i a_{j})/\sqrt{2}
\label{deij}
\end{eqnarray}
By gauge invariance imposed by the Gutzwiller projector,   $\langle \Delta_{s,d}\rangle=0$. However, $\Psi[u,v_{s (d)};x > 0]$ describe  
true $s$ ($d$)-wave superconductors as seen by the (off-diagonal) long range order in $\Delta_{s,d}$\cite{havilio-thesis}. 

In contrast, the insulator states  $\Psi[u,v_{ins},x]$ and the ``metallic'' states $\Psi[u,v_{met},x]$ have no long range superconducting order 
of either symmetry\cite{havilio-thesis}.

\begin{figure}
\input epsf
\centerline{ \epsfysize 7.0cm
\epsfbox{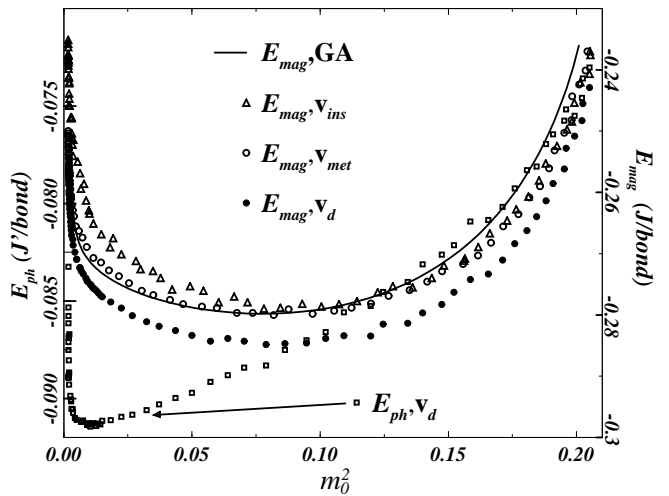}}
\caption{The  magnetic  energy  $E_{mag}$ (\protect{\ref{qhaf}}) 
and Cooper pair hopping energy $E_{ph}$ (\protect{\ref{jprime}})  versus local magnetization squared $m_0^2$, used as a variational parameter.
The  density of holes is 0.1 and lattice size is  $L=40$.
The magnetic energy is minimized at $m_0^2\approx 0.08$, consistent with a diluted quantum Heisenberg antiferromagnet, 
and is weakly dependent on inter-hole correlations. The points of $E_{mag}(v_s)$   overlaps  $E_{mag}(v_d)$. 
In contrast,  Cooper pair hopping prefers {\em vanishing} $m_0$
at $L \to \infty $. }
\label{sppemag}  	 	 
\end{figure}

{\em Effective Hamiltonians:}
Magnetic order is driven by the diluted Heisenberg model\cite{comm-stripes},
\begin{equation}
{\cal H}^{J}=J\sum_{\langle ij\rangle} {\bf S}_i \cdot {\bf S}_j (1-f^\dagger_i f_i)
(1- f^\dagger_j f_j)
\label{qhaf}
\end{equation}
where e.g.  $S^x+iS^y \equiv a^\dagger b$.
In Fig. \ref{sppemag} the  expectation value $E_{mag}(p)=\langle{\cal H}^{J}\rangle$  is plotted as a function of $m_0^2(p)$
for $x=0.1$  and various choices of 
$v$ from (\ref{uv}).
Within numerical errors all states minimize ${\cal H}^{J}$ at around 
$p_{min}\approx 2.7$,
which by Fig. \ref{sppvsp}(a) yields local magnetization of $m^2_0=0.08\approx m^2_0(0) (1-2x)$, where $m_0(0)$ 
agrees  with the  ground state local
magnetization of the undoped
Heisenberg model\cite{sandvik}. We have found that  $p_{min}\approx 2.7$ appears to
be independent of $x$ in for  $0 \le x\le 0.15$.
Thus we conclude that aside from the trivial kinematical
constraints, {\em the hole density and  correlations have little effect  
on the magnetic energy at low doping.}

A single hole hopping in the antiferromagnetic 
background has been shown by semiclassical
arguments\cite{t-JSC,Book}, to be effectively restricted at low energies
to hopping between sites on the same sublattice:
Next we consider the single hole hopping process
\begin{equation}
{\cal H}^{t'}= \sum_{\langle i k\rangle\in A,B}t_{ik}' f^\dagger_i f_k (a^\dagger_k a_i +b^\dagger_k b_i) 
\label{tprime}
\end{equation}
where $i,k$ are removed by two adjacent lattice steps, and $t' > 0$.
Unconstrained, the single hole ground state of $H^{t'}$ has momentum
on the edge of the magnetic Brillouin zone,  in agreement with
exact diagonalization of $t-J$ clusters\cite{dagotto}.
(Note: {\em inter}-sublattice  hopping (the t-term in the t-J model) is a high energy process in the
antiferromagnetic background. It primarily  renormalizes $t'$
and the quasiparticle's short distance structure\cite{t-JSC,Book}. 

The single holes hopping (\ref{tprime}) 
prefers the metallic state $v=v_{met}$ over the superconductor
$v=v_{s}, v_d$\cite{havilio-thesis}.  It also
prefers longer range $u(r)$ and thus actually {\em enhances} magnetic order
at low doping.  This is a type of
a Nagaoka effect, where mobile holes seperately polarize 
each of the  sublattices ferromagnetically.

Now we consider Cooper pairs hopping terms
\begin{equation}
{\cal H}^{J'}=-J'\left(\sum_{ijk}\Delta^{\dagger}_{ij}\Delta_{ik}+
\sum_{\langle i j\rangle, \tilde{i}\tilde{j}}\Delta^{\dagger}_{ij}\Delta_{\tilde{i}\tilde{j}}\right)
\label{jprime}
\end{equation}
The first term is derived from the large $U$ Hubbard model to order
$J' =t^2/U$\cite{Book}. It includes a {\em rotation} of the singlet pair,
which 
prefers $v_d$ over $v_s$. The second term is a parallel {\em translation}
of singlets. It  prefers superconductivity
with $v=v_d$ over metallic states with $v=v_{met}$\cite{comm-PH}.

In Fig. \ref{sppemag}  the ground state energy $E_{ph}$ of (\ref{jprime}) is
plotted for $v=v_{d}$, $x$=0.1 and $L$=40. For $v=v_s$, $E_{ph}>0$. 
The variational energy is minimized at  $p=3.35$, 
which by the finite size scaling of Fig.\ref{sppvsp}(b)
indicates vanishing $m_0$ at large $L$. Note the striking difference between 
the minima of $E_{mag}$ and $E_{ph}$. 
{\em Thus, Cooper pair hopping drives the groundstate toward a spin liquid phase!}
 
A simple explanation is that  pairs can hop with greater overlap when
parallel bonds have maximum {\em singlet} components. When $u(r)$ is longer ranged, 
triplet contributions are larger, which inhibits pair delocalization.
Incidentally,  the Gutzwiller approximation (GA) fails to predict this effect since
it decouples the local  correlations  between spins and hole
pairs.

Since ${\cal H}^{J'}$ is the effective model which
drives superconductivity it produces phase stiffness, which in
the continuum approximation is given by
\begin{equation}
{\cal H}^{J'}\approx {{V_0}\over{2}} \int  d^2 x (\nabla\phi_i)^2
\label{hphi}
\end{equation}
The stiffness constant $V_0$ can be determined variationally from the
RVB states, 
by imposing a uniform gauge field twist on
the bond parameters  $v_{i,j}\to v_{i,j} exp(i(x_i+x_j) \phi/2L)  $ and measuring  $E_{ph}(\phi)$ to find  $V_0=  d^2 E_{ph}/ d\phi^2$.

Following  Ref. \cite{emkivsfs}, at low doping for the square lattice  $V_0$ is roughly 
equal to $T_c$. 

\begin{figure}
\input epsf
\centerline{ \epsfysize 7.0cm
\epsfbox{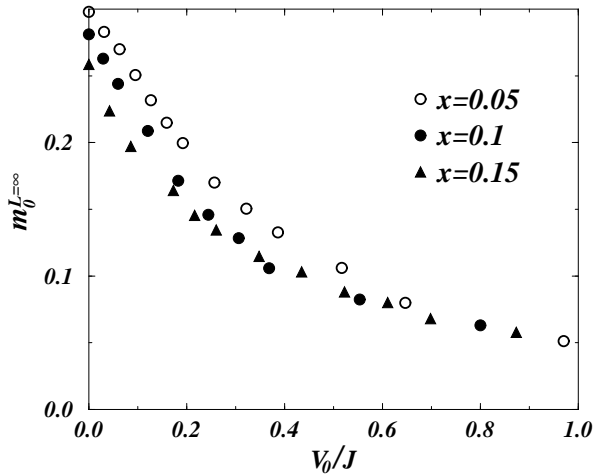}}
\caption{The relation between
thermodynamic local magnetization $m^{L=\infty}_0$ and superconducting  
phase stiffness $V_0$
(related to $T_c$, see text).
$J$ is the Heisenberg exchange energy.  The points are considered 
{\em upper bounds} on $m_0$, which may even vanish for $V_0/J \ge 0.2$.}
\label{jproverj}  	 	 
\end{figure}

{\em results}
In Fig.(\ref{jproverj}) we show our main result: The staggered magnetization $m_0$ 
for  ${\cal H}^{J}+{\cal H}^{J'}$ is plotted against the 
superconducting to magnetic stiffness ratio  $V_0/J$ for different doping concentrations $x=0.05,0.1,0.15$.

Two primary observations are made: (i) The local magnetization is 
sharply reduced at relatively low superconducting stiffness  (and  $T_c/J$). 
(ii) The relation between $m_0$ and $V_0/J$ appears to be  
weakly dependent
on the precise hole concentration. 

{\em Discussion:}
Because of finite size uncertainty, $m_0$ in Fig.(\ref{jproverj})  is
an {\em upper} bound on the thermodynamic local magnetization. The GA
extrapolation suggests that $m_0$ may actually vanish
already $V_0/J \ge 0.2$. This is in
qualitative agreement with the doping dependent of the local magnetization
measured  by Refs. \cite{Wakimoto,Niedermayer},  which diminishes
rapidly above the onset of superconductivity.

In a quantized theory of stripes\cite{stripes}, mechanisms for diminishing $m_0$ 
assume anisotropic magnetic couplings,
or fluctuating anti-phase domain walls. A direct connection between 
superconductivity and $m_0$ is not obvious in these approaches.

In a recent  {\em projected} SO(5) theory\cite{pSO5}, 
spins and hole-pairs dynamics have been considered with excluded
double occupancies. A variational relation is obtained
between superconducting stiffness and 
the magnetic order parameter,  which resembles the results of this paper.

{\em Acknowledgments:}
Useful conversations with A. Aharony, O. Entin, C. Henley, S. Kivelson
and  S-C. Zhang, are gratefully acknowledged.  MH thanks Taub computing center for support.  AA is supported by the Israel Science
Foundation  and the Fund for Promotion of Research at Technion.

\end{document}